# Polarization Dependent Loss and All-Optical Modulation in Graphene on Suspended Membrane Waveguides

Zhenzhou Cheng [1], Hon Ki Tsang [1, *], Xiaomu Wang [1], Chi Yan Wong [1], Xia Chen [2], Ke Xu [1], Zerui Shi,[1] and Jian-Bin Xu [1, *]

[1]Department of Electronic Engineering, The Chinese University of Hong Kong, Hong Kong

[2] Optoelectronics Research Centre, University of Southampton, United Kingdom.

We observe a strong polarization dependent optical loss of in-plane light propagation in silicon waveguide due to the presence of graphene. Both transverse-electric (TE) and transverse-magnetic (TM) modes are efficiently (~3 dB) coupled to the graphene on suspended membrane waveguides using an apodized focusing subwavelength grating. The TE mode has 7.7 dB less excess optical loss than the TM mode at 1.5 μm for a 150 μm long waveguide in good agreement with a theoretical model. All-optical modulation of light is demonstrated. There is also a large thermally induced change in waveguide effective index because of optical absorption in graphene.

[*] Electronic mail: hktsang@ee.cuhk.edu.hk and jbxu@ee.cuhk.edu.hk

Graphene has attracted much interest for optoelectronic applications because of its unique gapless band structure. [1] It has been used for saturable nonlinear absorption, [2,3] ultra-wideband absorption, [4,5] and tunable interband transition. [6,7] However, most of the previous experimental work employed light with normal incidence to the graphene layer and free space coupling, [2-7] which limits the light interaction length to the monolayer thickness (~0.7 nm). Only 2.3% optical absorption, predicted by fine structure constant, [8] can be observed in graphene for light of normal incidence, and 97.7% optical power will transmit without absorption, as shown in Fig. 1 (a). However, the two-dimensional (2D) nature of graphene makes it suitable for integration with planar lightwave circuits (PLCs), which can dramatically increase the interaction length between the graphene layer and light in the waveguide's evanescent field by many orders of magnitude. Moreover, it is possible to use complementary metal-oxide-semiconductor (CMOS) compatible fabrication processes to make the silicon waveguide devices, transfer or deposit the graphene onto the silicon [9,10] and pattern the graphene. [11,12] Hybrid graphene-silicon PLCs offer promising properties for future photonic devices. Experimental studies of graphene integrated PLC devices have recently been used for optical modulators, [13,14] and four-wave-mixing in silicon photonic crystal. [15] The optical losses in graphene on hydrogen silsesquioxane cladding silicon waveguide were also recently reported. [16]

Suspended membrane waveguide (SMW) devices have many attractive potential applications, such as energy efficient CMOS interconnects, [17] evanescent field sensors, [18] and optical force actuation. [19] The SMW can be fabricated on a silicon-on-insulator (SOI) wafer by locally etching the buried oxide (BOX) underneath the waveguide. The SMW devices were previously studied for mid-infrared (mid-IR) silicon photonics, as the removal of the BOX under the waveguide can avoid the large absorption losses of silica at mid-IR wavelengths. The integration of graphene and SMW, as shown in Fig.1 (b), can take full advantage of the transparent wavelength region of silicon and wide spectral window of graphene, which covers from near-IR to mid-IR. [20, 21] Moreover, with nanoscale high index contrast silicon SMW, the guided light can strongly interact with the surface graphene layer and reduce the device footprint. Graphene on SMWs can be operated over a broadband wavelength range from 1.2 μm to 8.0 μm. However, the optical interaction of light propagating in the plane of the graphene has not previously received much experimental attention.

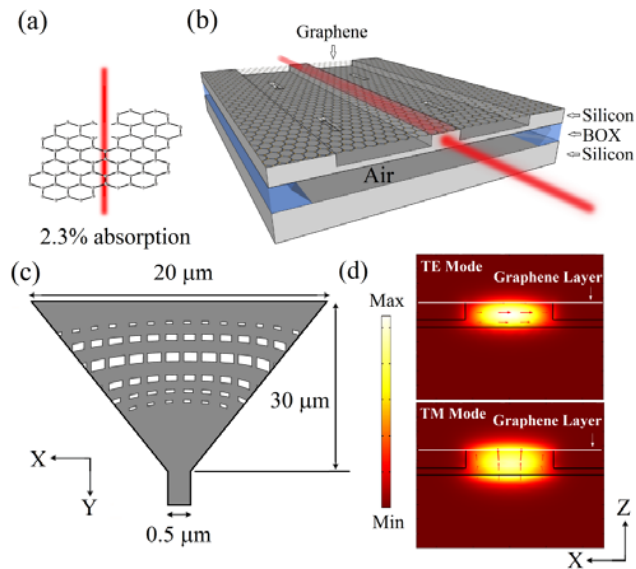

Fig.1 (a) Schematic picture of graphene absorption of normal incident light. (b) Schematic picture of in-plane light absorption in the graphene on SMWs. (c) Top view of apodized focusing SWG. (d) Finite element method (FEM) simulated electric field profile of $TE_{11}$ mode and $TM_{11}$ mode in the graphene on SMWs at 1.50 μm. The color map shows the optical intensity and arrows indicate the electric field.

In this paper, we present measurements of the polarization dependent optical loss of the graphene on SMWs on SOI. An apodized focusing subwavelength grating (SWG) was employed for high efficiency coupling of both transverse-electric (TE) and transverse-magnetic (TM) modes into the SMW. The optical losses for the TE and TM modes were characterized initially on a waveguide without graphene and on an identical waveguide with graphene. With only 150 μm length, the graphene on SMWs shows strong polarization dependent loss which agrees well with a theoretical model. All-optical modulation of light in the graphene on SMWs was also demonstrated. We also measured a large thermally induced change in the effective refractive index (RI) of the waveguide produced by the absorption of pump light.

The SMWs were designed and fabricated on a SOI wafer (supplied by SOITEC Inc.), which had 340 nm top silicon and 2.0 μm BOX. Periodical holes were etched beside the rib waveguide to enable the local removing of the BOX underneath the device. The grating couplers offer the advantages of allowing precise control of waveguide length without the need for cleaving or polishing of waveguide facets, [22] and high efficiency coupling [23] to standard single mode fibers (SMFs). The apodized focusing SWG, as shown in Fig.1 (c), was designed and fabricated for high efficiency coupling of both TE and TM modes into the SMW. The structure, including SMWs and gratings, was suspended in air, and thus can be used in both near-IR and mid-IR applications. The fabricated devices comprised a pair of nominally identical gratings connected with a 150 μm SMW. The silicon device design and fabrication process were described in the supplementary materials. The grating performance was characterized by a tunable laser, and a fiber polarization controller (FPC) was employed to change the polarization direction of the light from the SMF. The grating couplers had a maximum coupling efficiency of -3.0 dB and ~50 nm 3 dB bandwidth for TM mode. With the same grating, for TE mode, only half of coupling spectrum was measured because the center wavelength has shifted to ~1445 nm which was the limit of our laser tuning range. The maximum coupling efficiency was measured to be -4.0 dB for TE light, and ~120 nm 3 dB bandwidth was predicted from our design simulations. The measured spectral efficiency and wavelengths agree well with the 2D finite difference time domain (FDTD) simulations, as shown in Fig.2 (a). The polarization of input light can be determined from the different peak wavelengths and bandwidths of grating couplers for TM and TE modes.

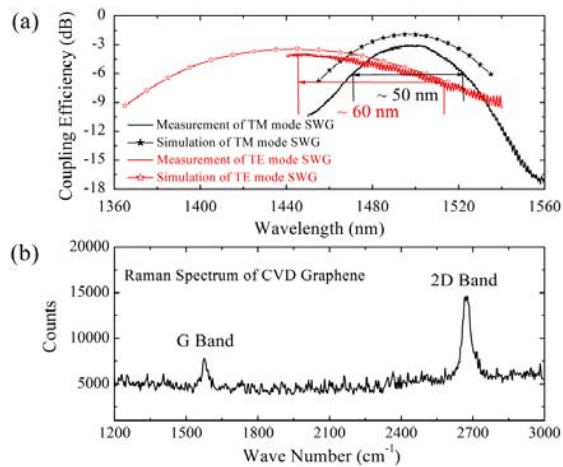

Fig.2 (a) FDTD simulations and measurements of apodized focusing SWGs for TE and TM modes without graphene. (b) Raman spectrum of CVD-grown graphene on the SMW which showing that the graphene film is monolayer.

The graphene sheets were grown on a copper foil by chemical vapor deposition (CVD). A thin PMMA layer was spin coated over the graphene layer and the copper was removed by wet etching. The graphene supported by PMMA was transferred to the SMW and the PMMA was then removed by acetone. Unlike the previous report,[13] the graphene was placed directly on the silicon waveguide without any planarization layer. As shown in Fig.2 (b), the monolayer graphene on the SMW was identified by the Raman spectroscopy via a symmetric 2D peak at 2671 cm$^{-1}$ with a full width at half maximum of 42.7 cm$^{-1}$ and G-to-2D peak intensity ratio (<0.5).[24] The scanning electron microscope (SEM) images of the graphene on SMWs and apodized focusing SWGs are shown in Fig.3. The SEM images had contrast between the regions with and without the graphene layer on silicon. It can be seen that there is some damage introduced during the graphene transfer process in Fig.3 (b) and Fig.3 (d).

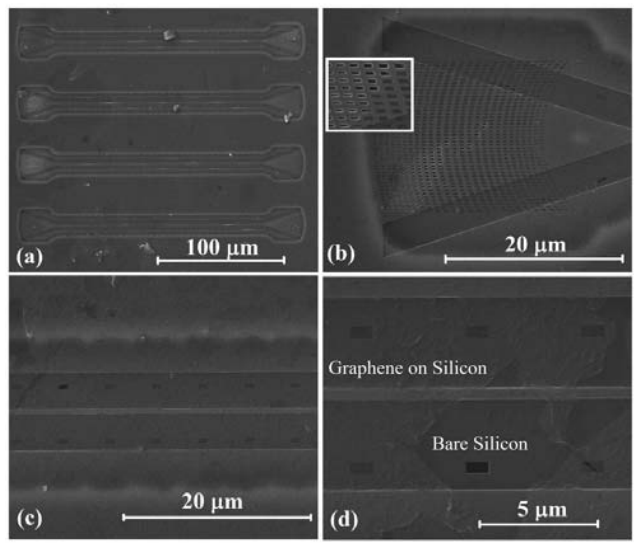

Fig.3 (a) SEM image of the graphene on SMWs and SWGs. (b) SEM image of apodized focusing SWG with the graphene on top. (c) SEM image of the graphene on SMW. (d) SEM image of the graphene on SMW with broken layer.

We fabricated four nominally identical gratings connected with 150 μm SMWs on two chips. One chip had the graphene layer transferred on top as described above. The fiber to fiber coupling losses of the chips were measured both for TE and TM modes at an input incident power of 0.1 mW. We observed a strong polarization dependent loss, as plotted in Fig.4. In our measurements, the incident light was coupled to SMWs at ~10 degrees from normal incidence, and thus we do not expect any graphene plasmon excitation to be present, [25] and the grating coupling efficiency is not affected significantly. Moreover, the coupling profile is almost unchanged for two chips in terms of bandwidth and center wavelength which indicates that the graphene layer also has a negligible influence on the RI of SWGs. Besides, we did not use any long adiabatic taper in the focusing grating coupler and thus avoided the possible mode conversion from TM mode to TE mode. [23] Thus the strong polarization dependent loss mainly comes from the graphene on SMWs.

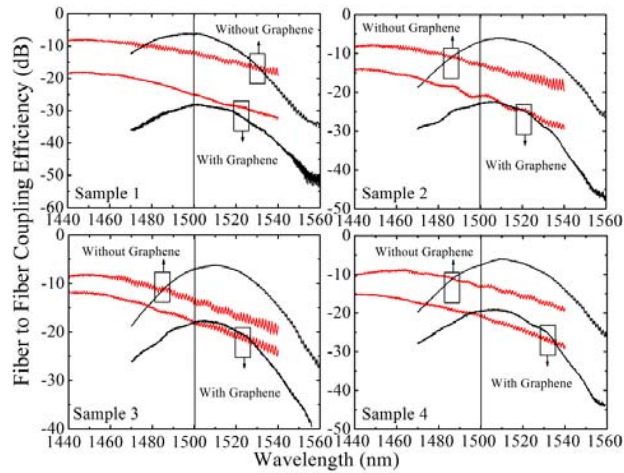

Fig.4 Measurements of fiber to fiber coupling efficiency both for TE and TM modes. Black (red) upper curves are for the TM (TE) mode without graphene and the black (red) curves are for the (TE) mode with graphene.

Four SMW samples, each of 150 μm length, were studied and the excess insertion losses introduced by the presence of the graphene layer were measured. Fiber to fiber excess losses of 10.1 dB, 6.1 dB, 3.4 dB and 5.0 dB for the TE mode were measured at its peak coupling wavelength (~1445 nm). While for TM mode, the measured excess losses introduced by the graphene were 22.4 dB, 16.9 dB, 12.0 dB and 13.2 dB, respectively at the peak TM coupling wavelength (~1510 nm). The different excess losses in the four samples mainly come from the discontinuity and non-uniformity of the graphene layer on SMWs. As the optical losses of TE and TM modes were recorded from the same waveguide, the defects/non-uniformity of graphene layer did not affect the measured differences in excess loss of the two polarizations. The excess loss differences of the two polarizations in the four samples measured were 7.7 dB, 6.1 dB, 7.1 dB and 6.4 dB, respectively, at 1.5 μm wavelength (indicated by the straight line in Figure 4). This polarization dependence is similar to the graphene-fiber polarizer in which a monolayer graphene was transferred onto 3.5 mm side-polished optical fiber.[26] To achieve the same excess loss (~27 dB) between two polarizations, the waveguide length in this work would need to be only ~500 μm instead of the 3.5 mm in the earlier work.[26] Due to the strong interaction between the graphene layer and evanescent field of silicon SMWs, the device footprint has been greatly reduced in this work. The difference in polarization dependent loss may be further enhanced by improving the graphene continuity and uniformity, and optimizing the top silicon thickness to increase the optical field strength at the graphene layer.

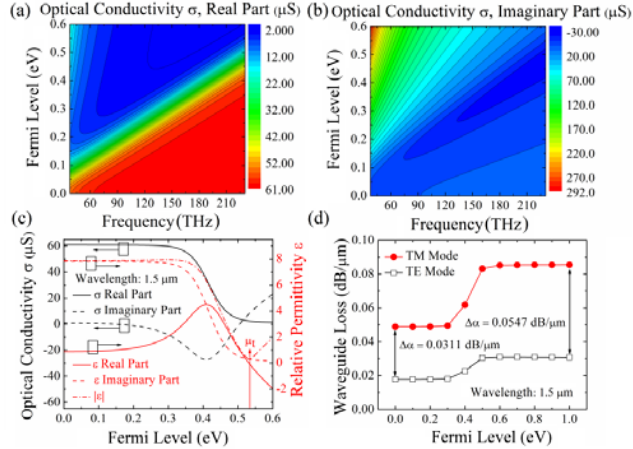

Fig.5 (a) Real part and (b) Imaginary part of the graphene optical conductivity as a function of the Fermi level and frequency (T = 300 K, $\Gamma$ = 5 meV), following the Kobo formula. (c) Graphene optical conductivity and effective relative permittivity as a function of Fermi level at 1.5 μm. (d) Theoretical optical loss of graphene on SMWs for TE and TM modes.

The imaginary part of optical conductivity plays an important role in the propagation light loss of the graphene on SMWs. The optical conductivity in this work was computed from the Kubo formalism, [27, 28] which is in agreement with the experiment results at temperature T = 3 K, with scattering rate $\Gamma$ = 0.43 eV. [7] The optical conductivity $\sigma$ of graphene includes the contributions from intraband and interband: $\sigma = \sigma_{intra}(\omega) + \sigma_{inter}(\omega)$. For intraband, the $\sigma_{intra}$ shows the following Drude-like form, [29] so intraband conductivity $\sigma_{intra}$ is always positive. While, for the $\sigma_{inter}$ can be positive or negative, [29] depending on the photon energy and Fermi level. When the interband contribution dominate, which gives the negative imaginary part, otherwise the $\sigma$ has positive imaginary part. The real part and imaginary part of the graphene optical conductivity as a function of the Fermi level and light frequency were calculated from near-IR to mid-IR, as shown in Fig.5 (a) and (b).

The effective relative permittivity of graphene can be determined by the following formula: [13]

$$\varepsilon_{eff} = 1 + i\frac{\sigma}{\omega\varepsilon_0\Delta} \qquad (1)$$

where the $\varepsilon_0$ is the vacuum permittivity, and $\Delta$ is the graphene effective thickness which equals to 0.7 nm in the simulations. [13] At 1.5 μm wavelength, the optical conductivity and effective relative permittivity were calculated as shown in Fig.5 (c). When the real part of relative permittivity is positive, the TE mode light can be guided in the graphene-silicon layer with weak damping, [30] and TM mode propagation light cannot be supported. However, for negative relative permittivity, the TM mode can excite the surface-plasmon polariton wave on the surface of graphene-silicon layer. So, at the Fermi level $\mu_t$, the graphene transforms from the "dielectric-like" graphene to "metallic-like" graphene.

For the graphene on SMWs, the light is mainly confined by the RI step between silicon and air, with a graphene layer perturbation. Based on above theoretical parameters, the $TE_{11}$ and $TM_{11}$ modes of graphene on SMWs were simulated at 1.5 μm wavelength using effective index method by FEM software as shown in Fig.1 (d) and Fig.5 (d). The attenuation of TM mode is obviously larger than that of TE mode, and the Fermi level will have larger influence on TM mode loss. The loss difference $\Delta\alpha$ between TM and TE modes were simulated to be 0.0311 dB/μm and 0.0547 dB/μm for "dielectric-like" graphene and "metallic-like" graphene, respectively. So, at 1.5 μm wavelength, with 150 μm length waveguide, the theoretical difference in excess loss of the two polarizations is predicted to be between 4.7 dB ("dielectric-like" graphene) to 8.2 dB ("metallic-like" graphene) which depends on the graphene Fermi level. The intrinsic doping level for our devices (the same batch of graphene sample on a similar substrate) is estimated as ~70 meV (or doping level about $1.2 \times 10^{12}$ /cm$^2$), so it is safe to absorb the 1.5 um laser. And the absorption of light will increase the free carrier population, and thus raise the quasi-Fermi level, and make the sample more "metallic-like". This theoretical prediction agrees well with the experimental results. Using this model but with the wafer of 220 nm-thick top silicon, the excess loss can be further increased to ~29 dB in a 150 μm length waveguide. In mid-IR wavelength range, the simulation predicts larger polarization dependent loss, due to the increased evanescent field that interacts with the graphene layer, and lower photon energy which would be more sensitive to the Fermi level.

The modulation of dielectric constant of graphene by an applied gate voltage has been studied previously. [13, 14] Here, we study the optical control of the Fermi level and use an intense pump light to modulate the optical absorption of a probe signal in the graphene on SMW. Furthermore the photo-excited hot carriers in graphene will produce a large change in the effective RI ($n_{eff}$) of SMWs. A high power (pump) laser (1439.2 μm) was coupled with TE mode polarization to control the Fermi level of graphene on SMWs, and transmission of TE mode and TM mode probes were measured at different pump powers.

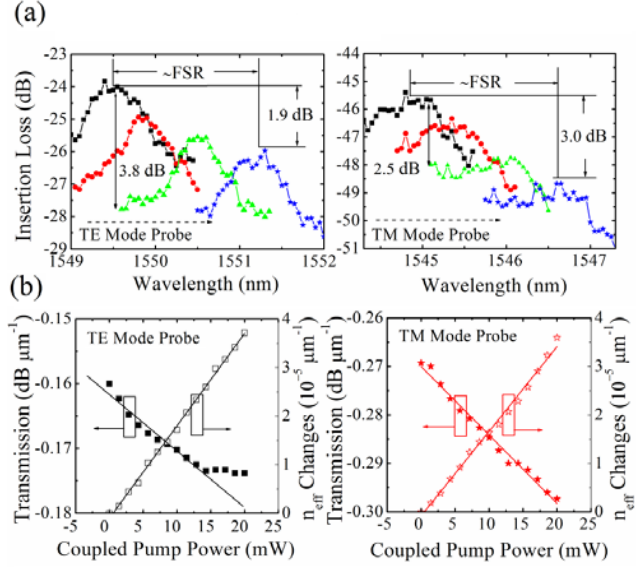

Fig.6 (a) Measurements of insertion loss under the control of TE mode pump power with optically induced shift of one FSR. The dash lines indicate the F-P shift with increasing the pump power. For TE mode probe, four coupled pump powers are 0 mW, 4.3 mW, 10 mW, and 15.7 mW, respectively. For TM mode probe, four coupled pump powers are 0 mW, 5.7 mW, 11.5 mW, and 17.2 mW, respectively. (b) The device transmission and the measured changes of effective RI with optical controlled Fermi level at different coupled pump powers.

The results in Fig.6 (a) demonstrate the all-optical modulation of the probe transmission by varying the pump power. The Fabry-Perot (F-P) oscillations in the probe transmission allowed us to obtain the changes in the real and imaginary parts of the $n_{eff}$ of the graphene on SMWs. As a control experiment we also performed the same measurement on a SMW without any graphene. For the control experiment, we observed only a small (0.6 dB) intensity dependence that may be attributed to the free carriers generated by two photon absorption in the 150 μm long silicon waveguide. However, for the graphene on SMWs, much larger intensity dependence (for TE mode, 3.8 dB loss at 1549.5 nm and 1.9 dB at peak wavelength) was observed. In Fig.6 (a), the transmission spectrum of the probe light has an F-P oscillation, with a free spectral range (FSR) of ~1.5 nm, because of the grating back reflection. Fig.6 (a) shows that the F-P oscillations shifted to longer wavelength with increasing pump power. Optical absorption in graphene will change the quasi-Fermi level position which modifies the graphene from a "dielectric-like" material to a more "metallic-like" material. From the theoretical model plotted in Fig.5 (d), the excess insertion loss introduced to the TM mode probe should be more than that of the TE mode probe, and the TE mode probe will saturate before the TM mode as the Fermi level is increased. These predictions are in agreement with our experimental observation. From Fig.6 (b), we observed loss saturation for TE mode, and the TM mode loss also start to saturate when the coupled pump power increased beyond 20 mW. The optical modulation coefficients were $8.72 \times 10^{-4}$ dB/μm/mW and $14.1 \times 10^{-4}$ dB/μm/mW for TE mode and TM mode probes, respectively. Another interesting phenomenon is the red-shift of F-P oscillations with increasing pump power. The observed shifts in F-P oscillations imply the presence of a large linear change of $n_{eff}$, as plotted in Fig.6 (b). The large changes in $n_{eff}$ are consistent with thermal-optic effects. The absorption of the pump power dramatically increases the hot carriers' density in the graphene; in the steady state, the relaxation of the hot carriers will eventually produce a temperature rise in the silicon membrane since the air cladding of SMW provides a good thermal isolation. In the experiment, the graphene layer suffered irreversible thermal damage ("laser burning") when the coupled power exceeded ~30 mW. From the measured F-P shifts, the temperature rise of SMW was calculated to be ~20.7 K with pump coupled power of 15.7 mW, which introduces a shift of about one FSR (~π phase shift). The all-optical modulation in graphene is intrinsically broadband as the overall optical opacity of graphene is independent of wavelength, and the high frequency dynamic conductivity for Dirac fermions is almost independent of wavelength. [31] The modulation depth can be further enhanced by optimizing the top silicon thickness. The results point to many potential applications of graphene on SMWs in the near-IR and mid-IR region, such as on-chip optically-switchable polarizer, efficient mid-infrared photodiodes and ultrafast all-optical logic gates.

In summary, we report measurements of the polarization dependent optical loss of graphene on SMWs. The TE and TM modes light are coupled into the waveguide with an apodized focusing SWG. The graphene clearly introduce higher excess losses for TM mode with a difference as large as 7.7 dB observed at 1.5 μm wavelength in a 150 μm length waveguide. We demonstrate all-optical modulation with measured optical modulation coefficients of $8.72 \times 10^{-4}$ dB/μm/mW and $14.1 \times 10^{-4}$ dB/μm/mW for TE mode and TM mode probes using a TE mode optical pump, respectively. The absorption of optical power also produced large thermal changes in the effective RI of the graphene on SMW samples.

Acknowledgments: We thank Mr. Yubin Xiao for preparing graphene samples. This work is supported by University Grants Committee special equipment grant SEG CUHK-01 and Hong Kong Research Grants Council GRF grants (Nos. 416512 and 417910). Zhenzhou Cheng would like to thank the support of the Research Grants Council Ph.D. Fellowship.

**References:**

[1] F. Bonaccorso, Z. Sun, T. Hasan, and A. C. Ferrari, Nat. Photon. **4**, 611 (2010).

[2] Q. Bao, H. Zhang, Y. Wang, Z. Ni, Y. Yan, Z. X. Shen, K. P. Loh, and D. Y. Tang, Advanced Functional Materials **19**, 3077 (2009).

[3] Z. Sun, T. Hasan, F. Torrisi, D. Popa, G. Privitera, F. Wang, F. Bonaccorso, D. M. Basko, and A. C. Ferrari, ACS Nano **4**, 803 (2010).

[4] A. B. Kuzmenko, E. V. Heumen, F. Carbone, and D. V. D. Marel, Phys. Rev. Lett. **100**, 117401 (2008).

[5] T. Stauber, N. M. R. Peres, and A. K. Geim, Phys. Rev. B **78**, 085432 (2008).

[6] F. Wang, Y. Zhang, C. Tian, C. Girit, A. Zettl, M. Crommie, and Y. R. Shen, Science **320**, 206 (2008).

[7] Z. Q. Li, E. A. Henriksen, Z. Jiang, Z. Hao, M. C. Martin, P. Kim, H. L. Stormer, D. N. Basov, Nat. Phys. **4**, 532 (2008).

[8] R. R. Nair, P. Blake, A. N. Grigorenko, K. S. Novoselov, T. J. Booth, T. Stauber, N. M. R. Peres, and A. K. Geim, Science **320**, 1308 (2008).

[9] G. Hong, Q. Wu, J. Ren, and S Lee, Appl. Phys. Lett. **100**, 231604 (2012).

[10] J. Mao, L. Huang, Y. Pan, M. Gao, J. He, H. Zhou, H. Guo, Y. Tian, Q. Zou, L. Zhang, H. Zhang, Y. Wang, S. Du, X. Zhou, A. H. C. Neto, and H. Gao, Appl. Phys. Lett. **100**, 093101 (2012).

[11] K. S. Novoselov, A. K. Geim, S. V. Morozov, D. Jiang, M. I. Katsnelson, I. V. Grigorieva, S. V. Dubonos, and A. A. Firsov, Nature **438**, 197 (2005).

[12] M. Y. Han, B. Özyilmaz, Y. Zhang, and P. Kim, Phys. Rev. Lett. **98**, 206805 (2007).

[13] M. Liu, X. Yin, E. Ulin-Avila, G. Geng, T. Zentgraf. L. Ju, F. Wang, and X. Zhang, Nature **474,** 64 (2011).

[14] M. Liu, X. Yin, and X. Zhang, Nano Lett. **12**, 1482 (2012).

[15] T. Gu, N. Petrone, J. F. McMillan, A. V. D. Zande, M. Yu, G. Q. Lo, D. L. Kwong, J. Hone, and C. W. Wong, Nat. Photon.**6**, 554 (2012).

[16] H. Li, Y. Anugrah, S. J. Koester, and M. Li, Appl. Phys. Lett. **101**, 111110 (2012).

[17] J. S. Orcutt, A. Khilo, C. W. Holzwarth, M. A. Popović, H. Li, J. Sun, T. Bonifield, R. Hollingsworth, F. X. Kärtner, H. I. Smith, V. Stojanović, and R. J. Ram, Opt. Express **19**, 2335 (2012).

[18] J. Hu, and D. Dai, IEEE Photon. Technol. Lett. **23**, 842 (2011).

[19] D. V. Thourhout, and J. Roels, Nat. Photon. **4**, 211 (2010).

[20] S. J. Koester, and M. Li, Appl. Phys. Lett. **100**, 171107 (2012).

[21] Z. Cheng, X. Chen, C. Y. Wong, K. Xu, and H. K. Tsang, IEEE Photon. J. **4,** 1510 (2012).

[22] S. Scheerlinck, D. Taillaert, D. V. Thourhout, and R. Baets, Appl. Phys. Lett. **92**, 131101 (2008).

[23] Z. Cheng, X. Chen, C. Y. Wong, K. Xu, and H. K. Tsang, Appl. Phys. Lett. **101**, 101104 (2012).

[25] X. Li, Y. Zhu, W. Cai, J. An, S. Kim, J. Nah, D. Yang, R. Piner, A. Velamakanni, I. Jung, E. Tutuc, S. K. Banerjee, L. Colombo, and R. S. Ruoff, Science **324**, 1312 (2009).

[26] T. R. Zhang, F. Y. Zhao, X. H. Liu, and J. Zi, Phys. Rev. B **86**, 165416 (2012).

[27] Q. Bao, H. Zhang, B. Wang, Z. Ni, C. H. Y. X. Lim, Y. Wang, D. Y. Tang, and K. P. Loh, Nat. Photon. **5**, 411 (2011).

[28] V. P. Gusynin, S. G. Sharapov, and J. P. Carbotte, J. Phys.: Condens. Matter **19**, 026222 (2007).

[29] Z. Lu, and W. J. Zhao, Opt. Soc. Am. B **29**, 1490 (2012).

[30] S. A. Mikhailov, and K. Ziegler, Phys. Rev. Lett. **99**, 016803 (2007).

[31] G. W. Hanson, J. of Appl. Phys. **103**, 064302 (2008).

[32] K. F. Mak, M. Y. Sfeir, Y. Wu, C. H. Lui, J. A. Misewich, T. F. Heinz, Phys. Rev. Lett. **101**, 196405 (2008).

**Figures Captions**

Fig.1 (a) Schematic picture of graphene absorption of normal incident light. (b) Schematic picture of in-plane light absorption in the graphene on SMWs. (c) Top view of apodized focusing SWG. (d) Finite element method (FEM) simulated electric field profile of $TE_{11}$ mode and $TM_{11}$ mode in the graphene on SMWs at 1.50 μm. The color map shows the optical intensity and arrows indicate the electric field.

Fig.2 (a) FDTD simulations and measurements of apodized focusing SWGs for TE and TM modes without graphene. (b) Raman spectrum of CVD-grown graphene on the SMW which showing that the graphene film is monolayer.

Fig.3 (a) SEM image of the graphene on SMWs and SWGs. (b) SEM image of apodized focusing SWG with the graphene on top. (c) SEM image of the graphene on SMW. (d) SEM image of the graphene on SMW with broken layer.

Fig.4 Measurements of fiber to fiber coupling efficiency both for TE and TM modes. Black (red) upper curves are for the TM (TE) mode without graphene and the black (red) curves are for the (TE) mode with graphene.

Fig.5 (a) Real part and (b) Imaginary part of the graphene optical conductivity as a function of the Fermi level and frequency (T = 300 K, $\Gamma$ = 5 meV), following the Kobo formula. (c) Graphene optical conductivity and effective relative permittivity as a function of Fermi level at 1.5 μm. (d) Theoretical optical loss of graphene on SMWs for TE and TM modes.

Fig.6 (a) Measurements of insertion loss under the control of TE mode pump power with optically induced shift of one FSR. The dash lines indicate the F-P shift with increasing the pump power. For TE mode probe, four coupled pump powers are 0 mW, 4.3 mW, 10 mW, and 15.7 mW, respectively. For TM mode probe, four coupled pump powers are 0 mW, 5.7 mW, 11.5 mW, and 17.2 mW, respectively. (b) The device transmission and the measured changes of effective RI with optical controlled Fermi level at different coupled pump powers.

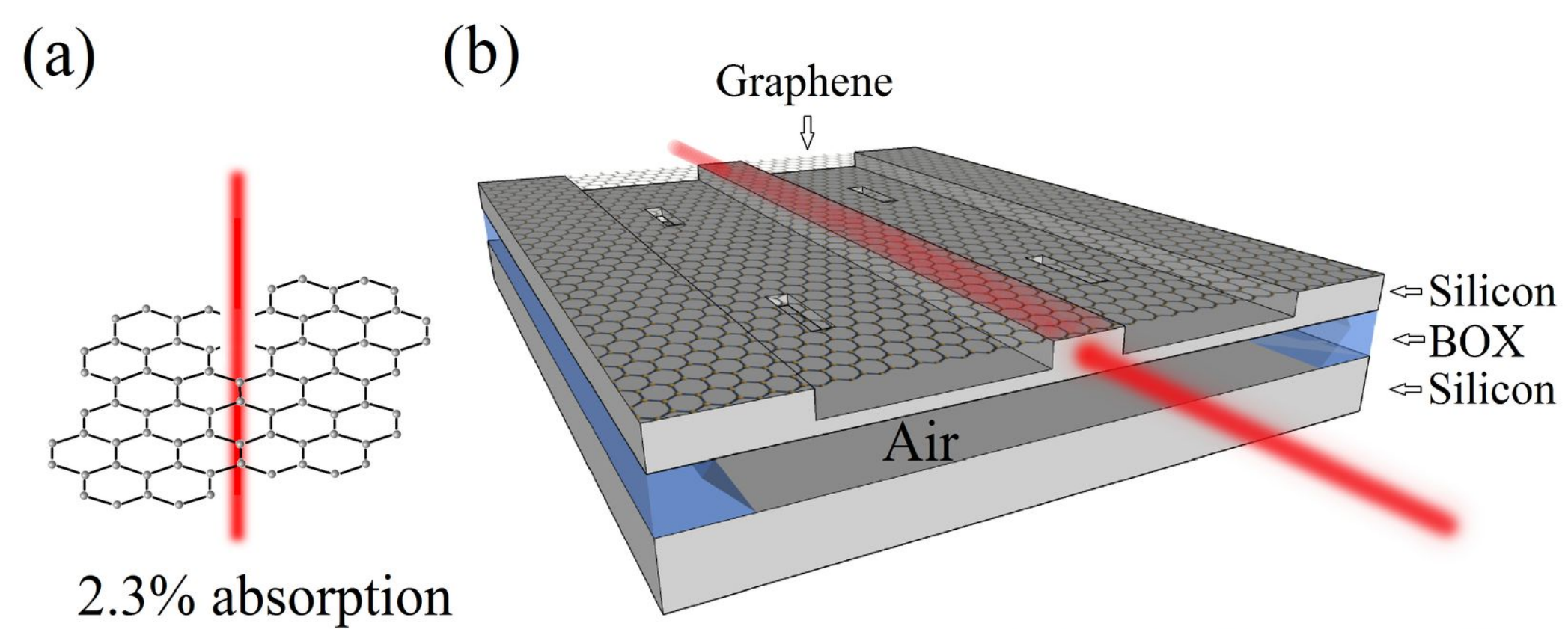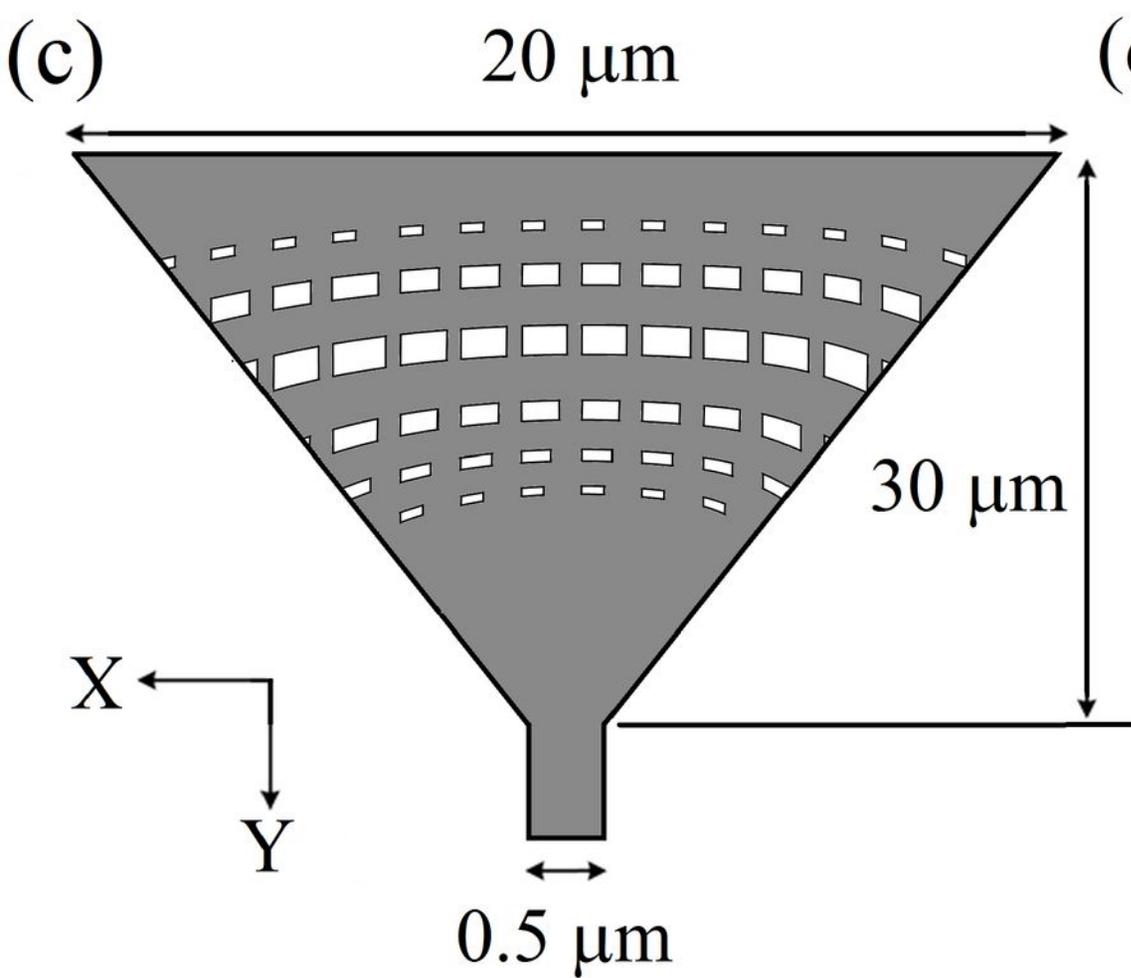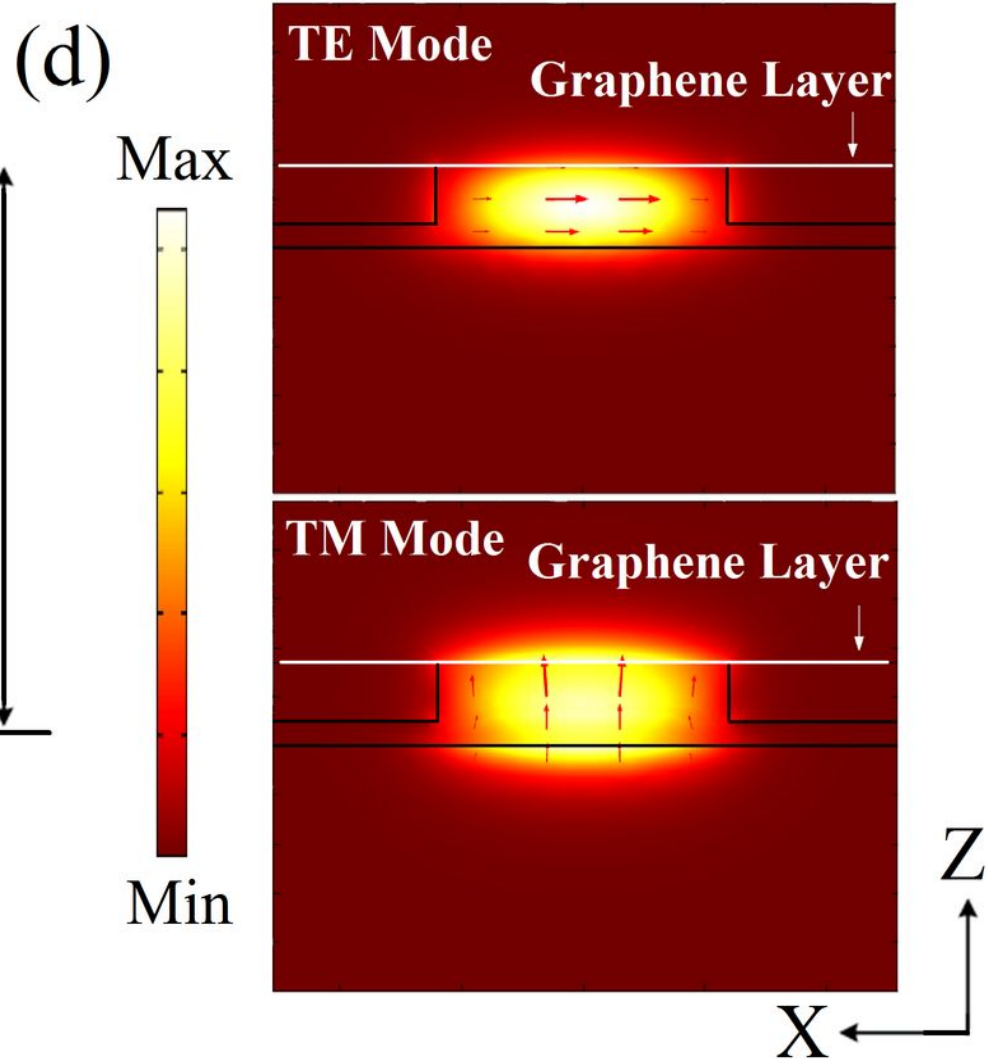

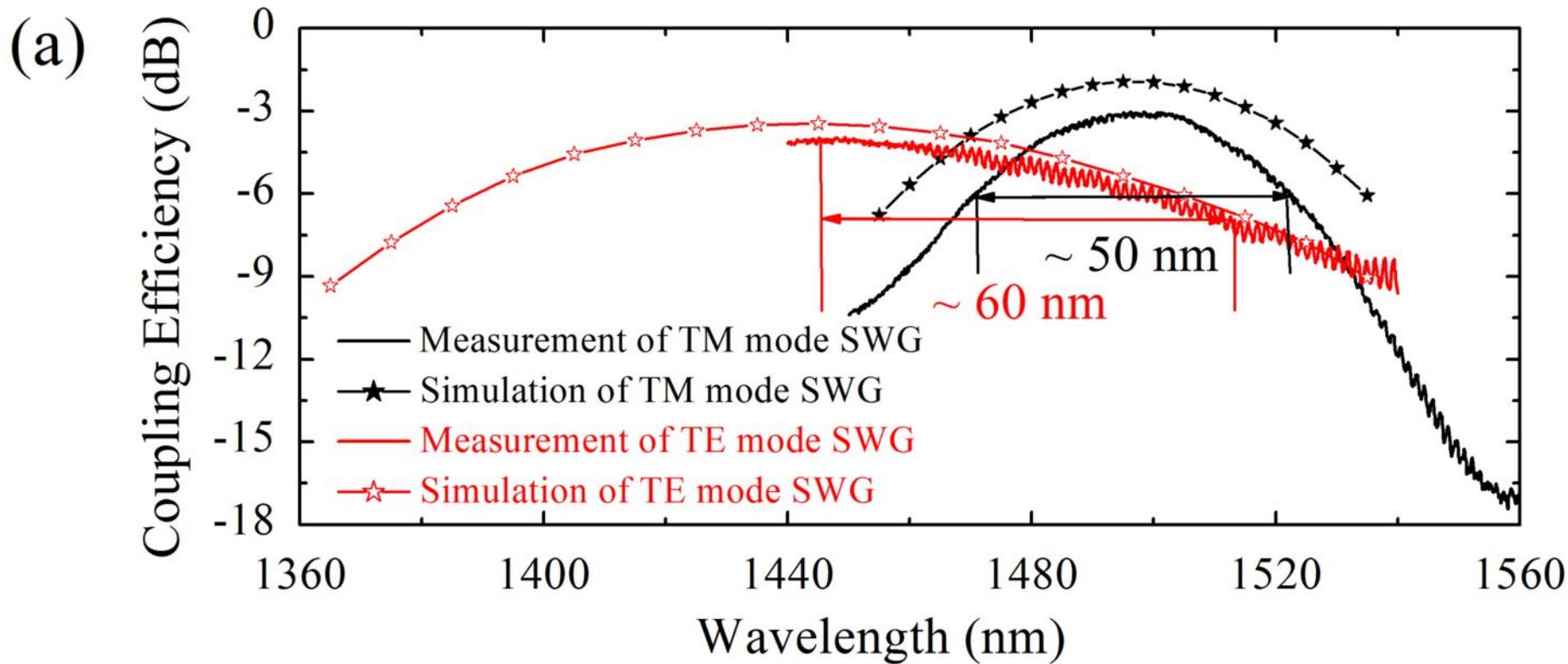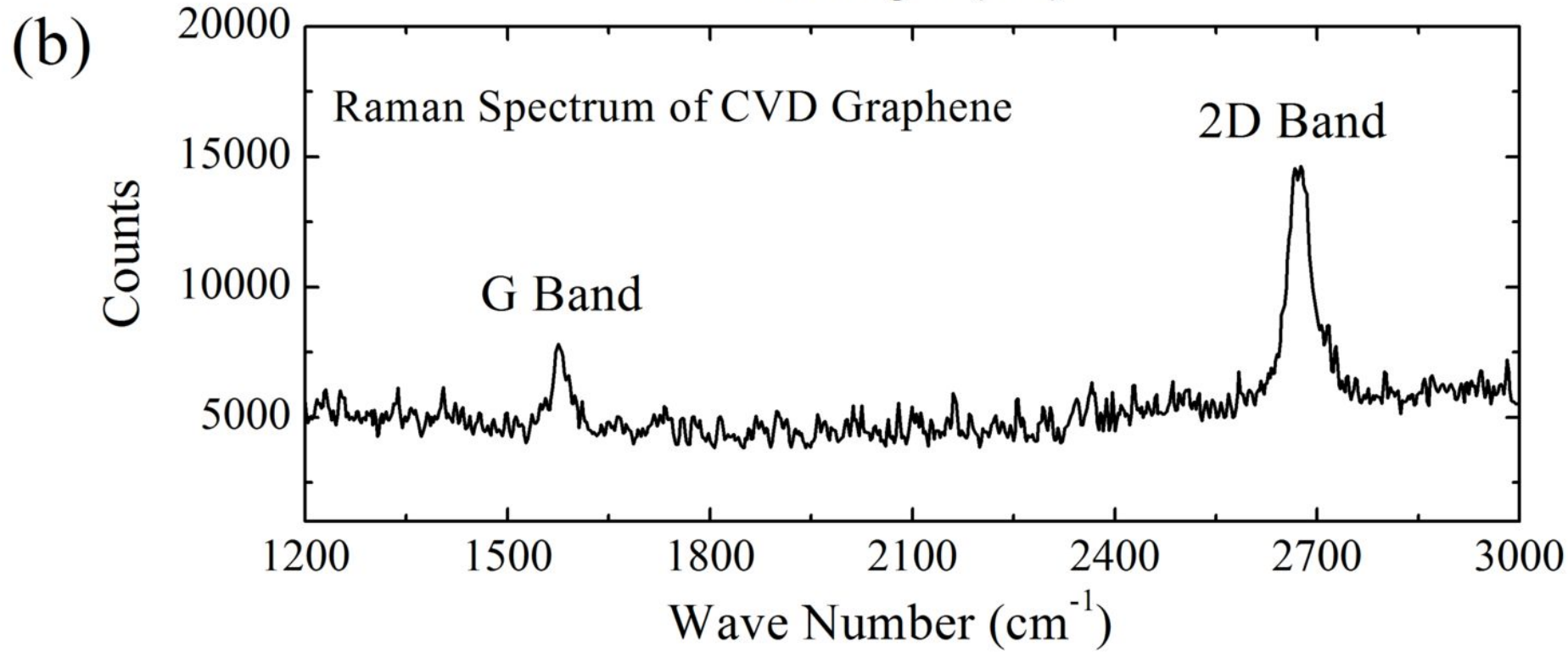

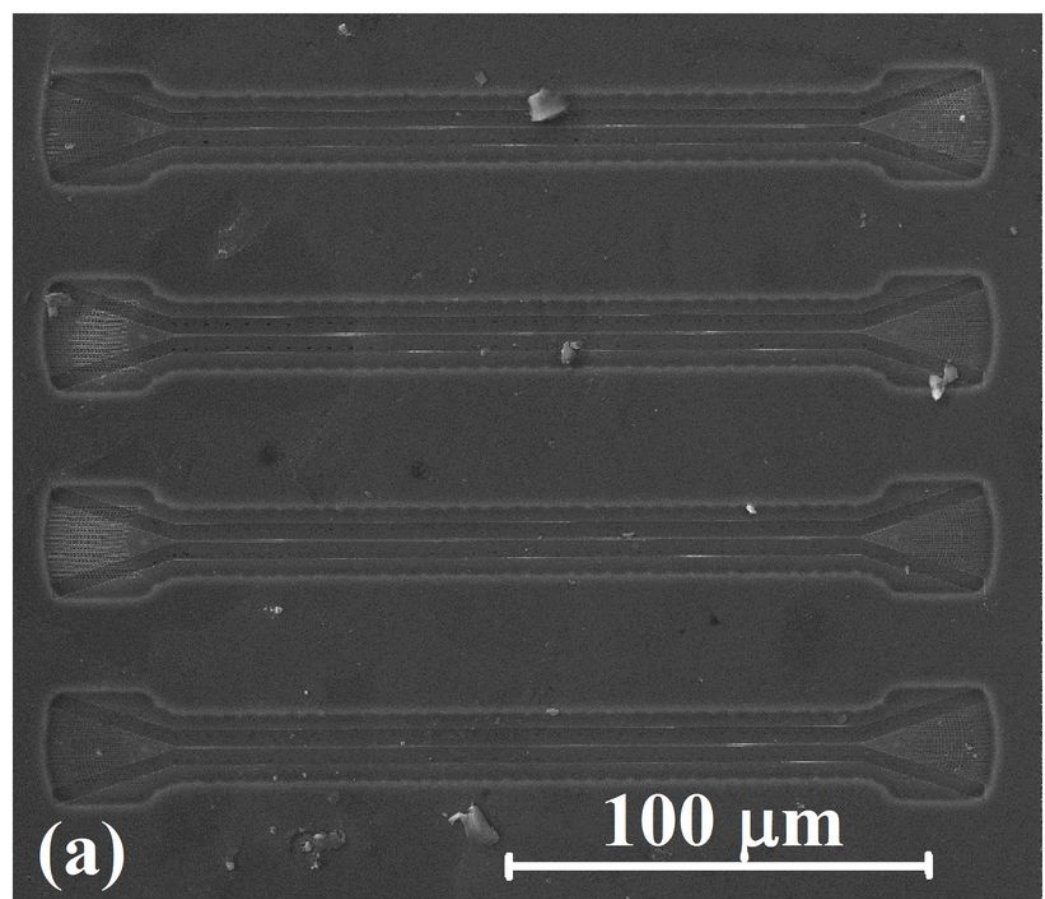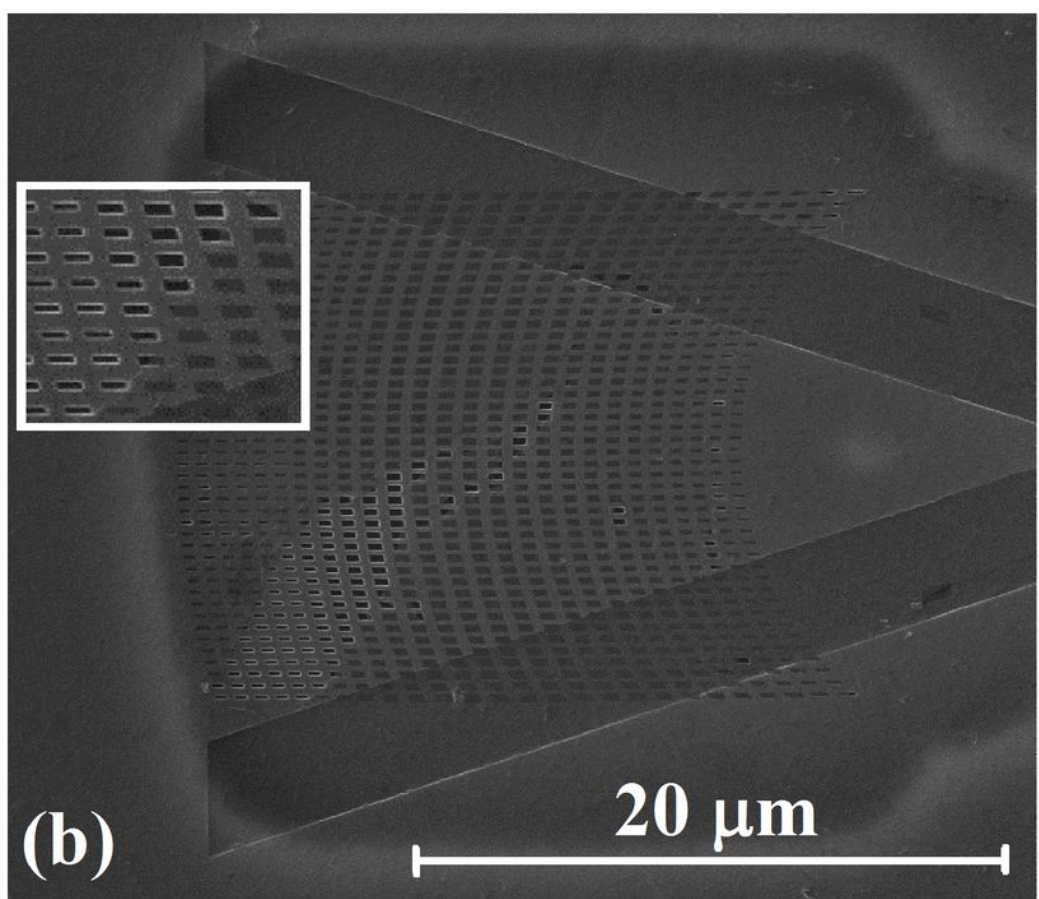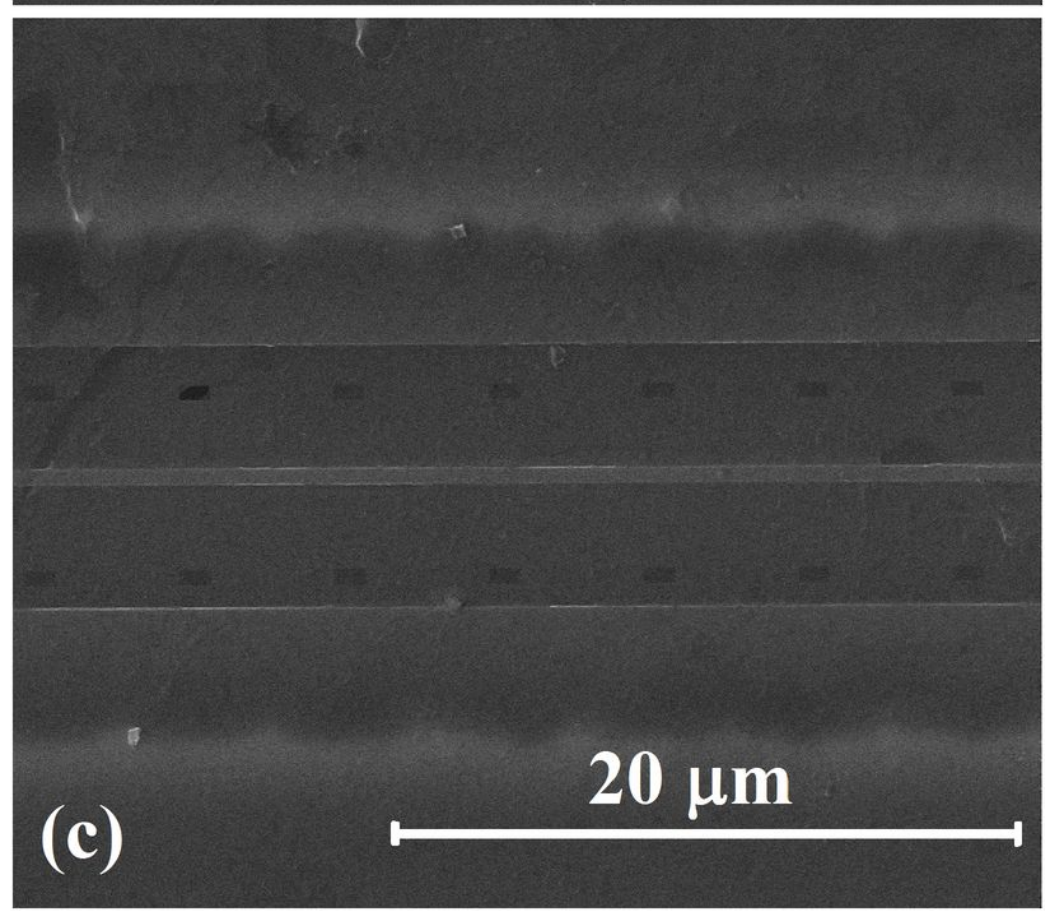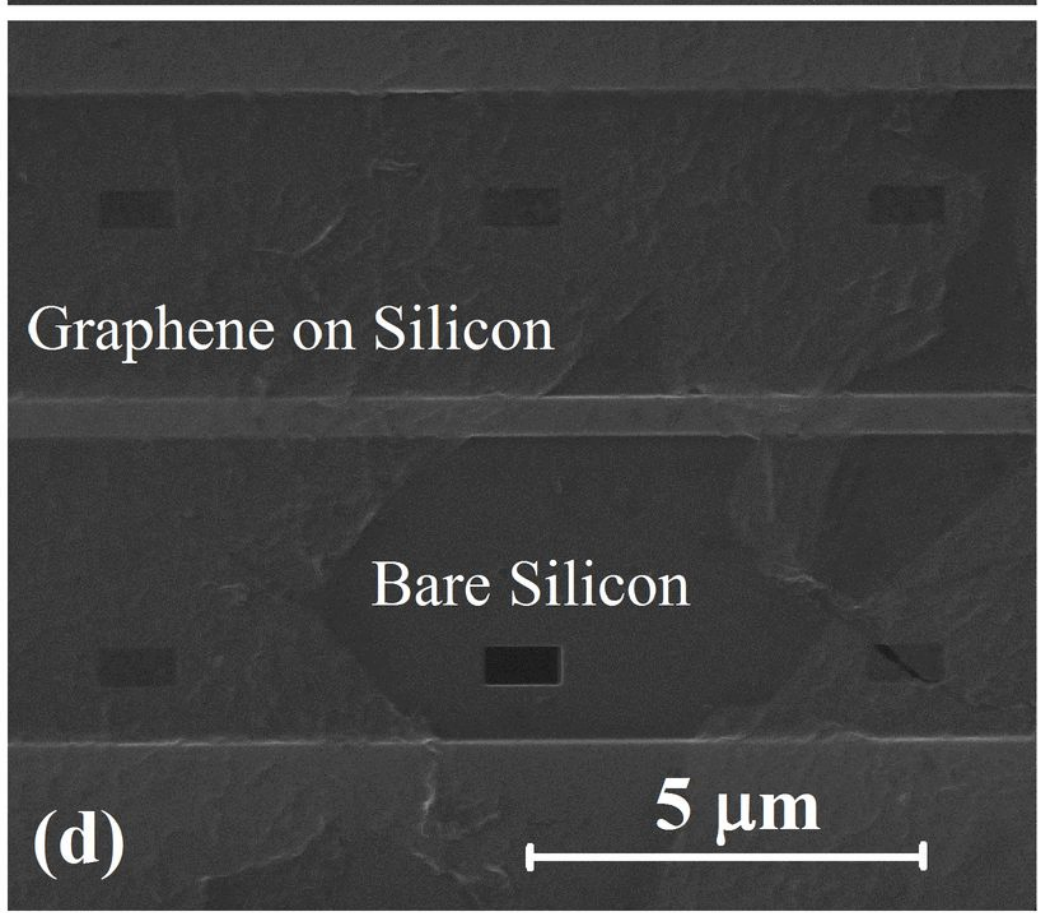

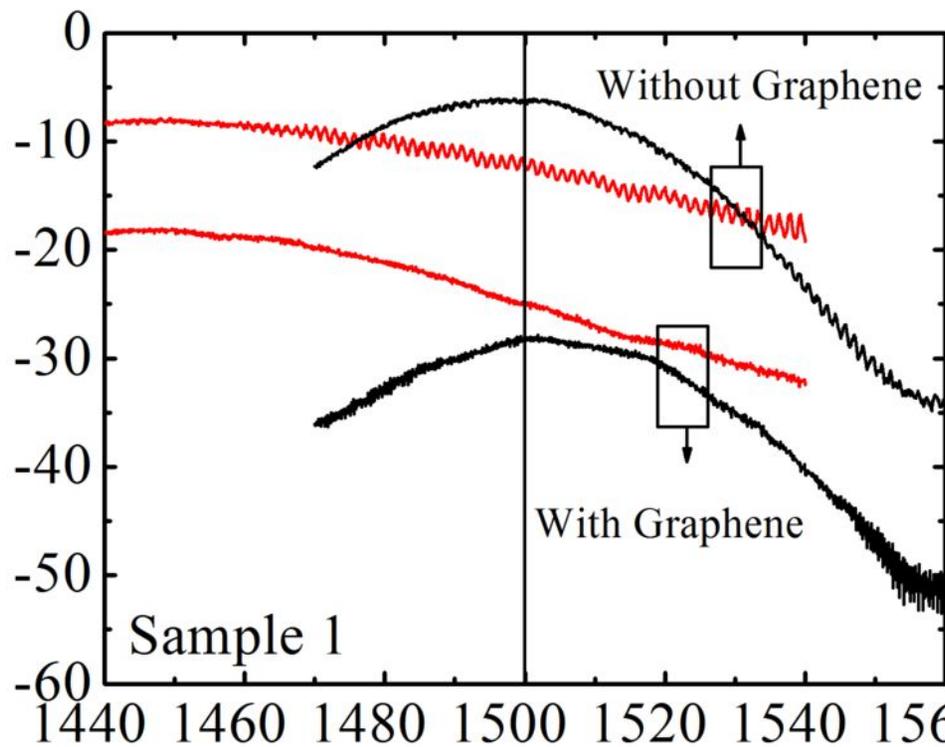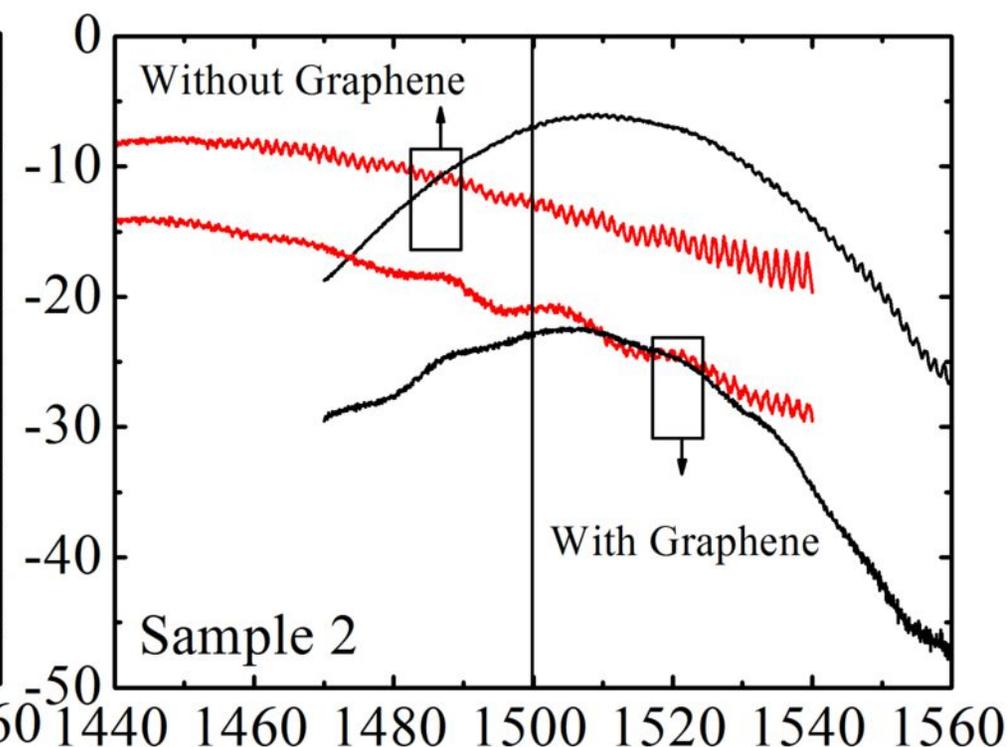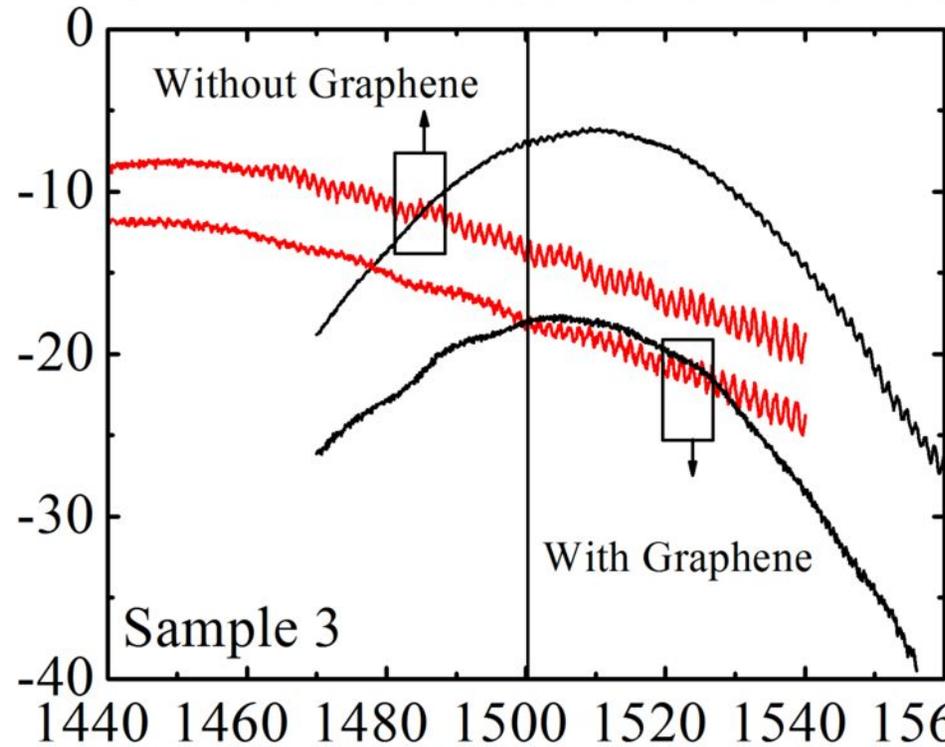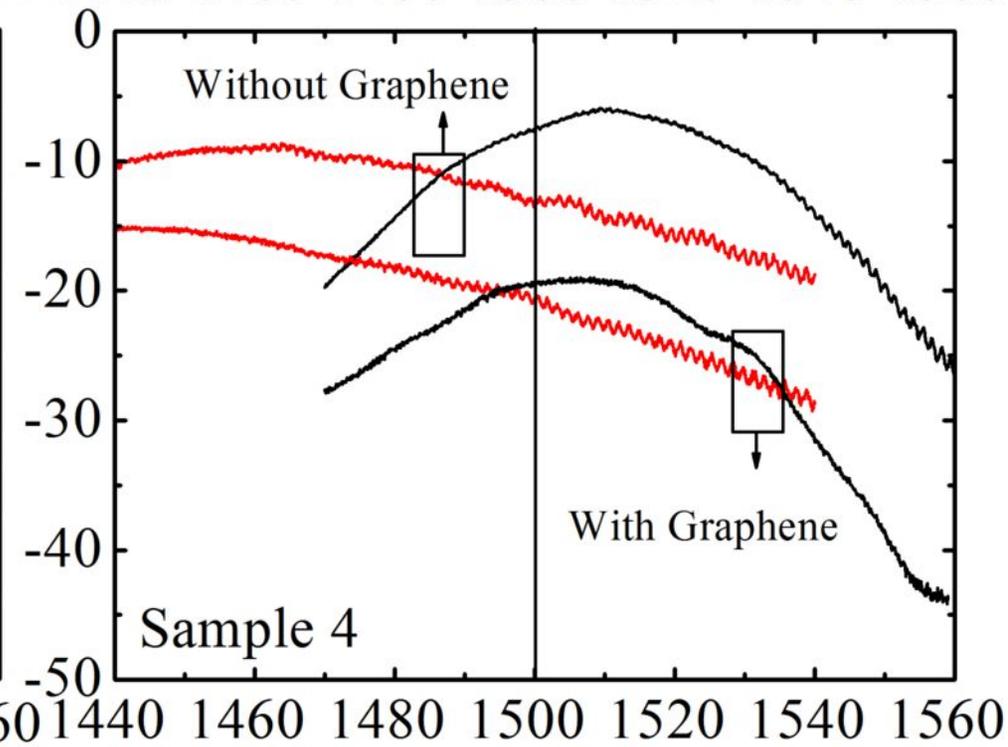

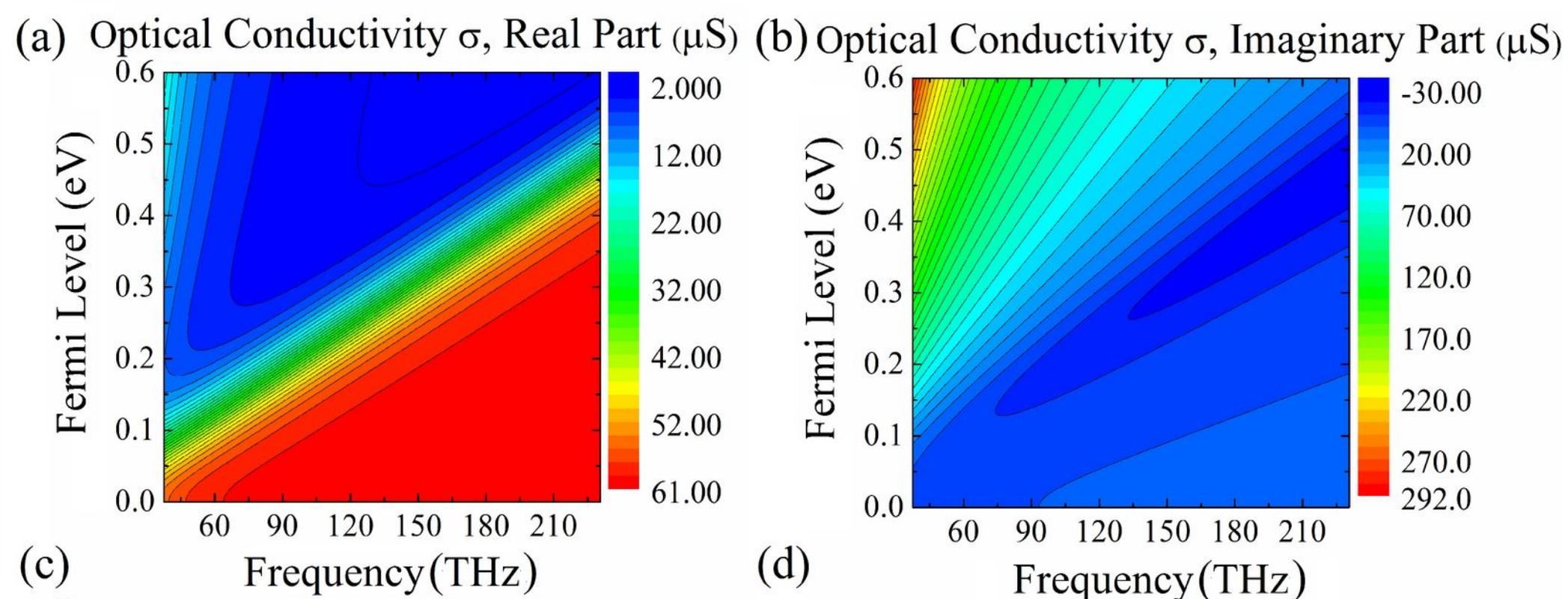

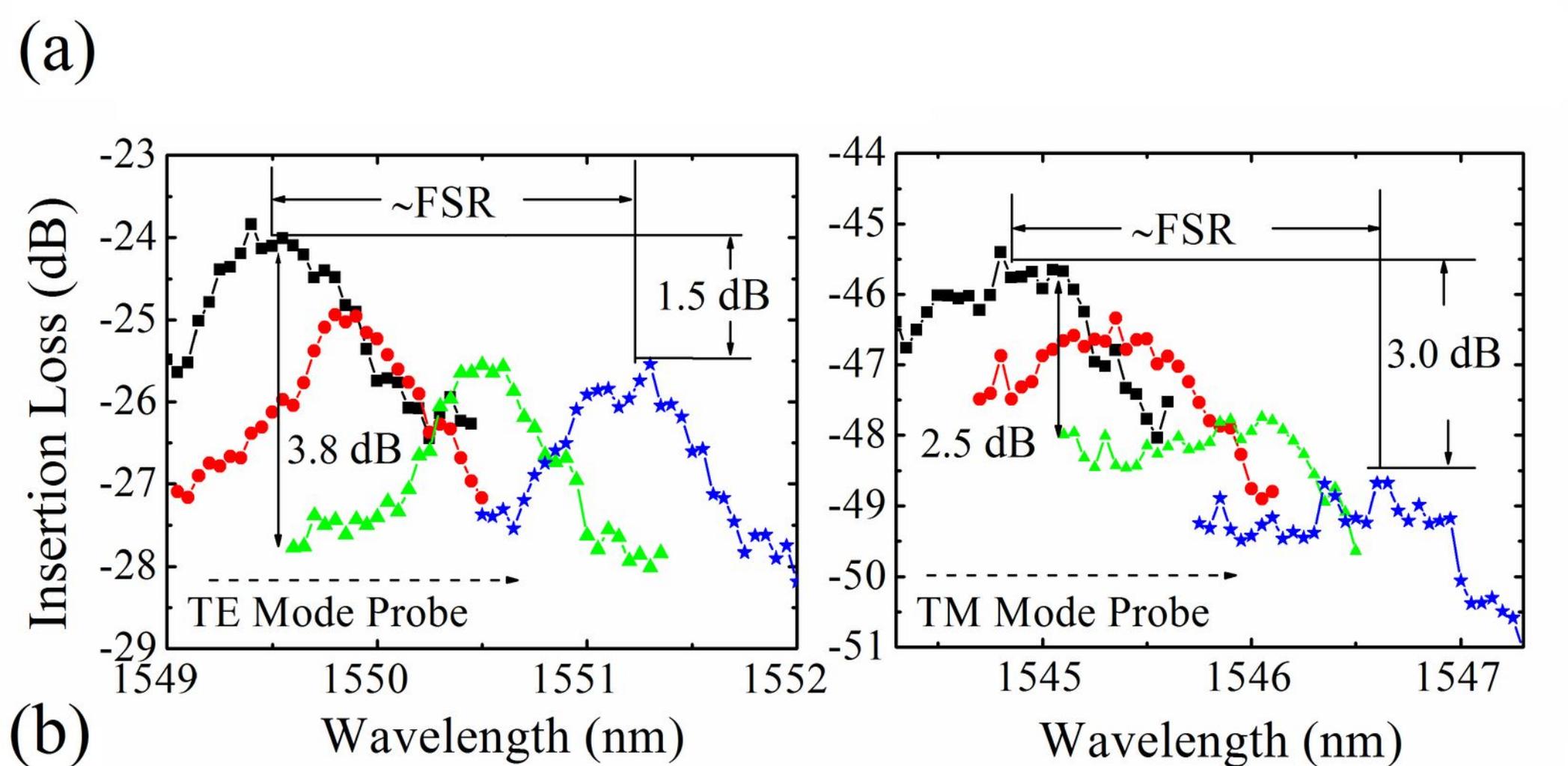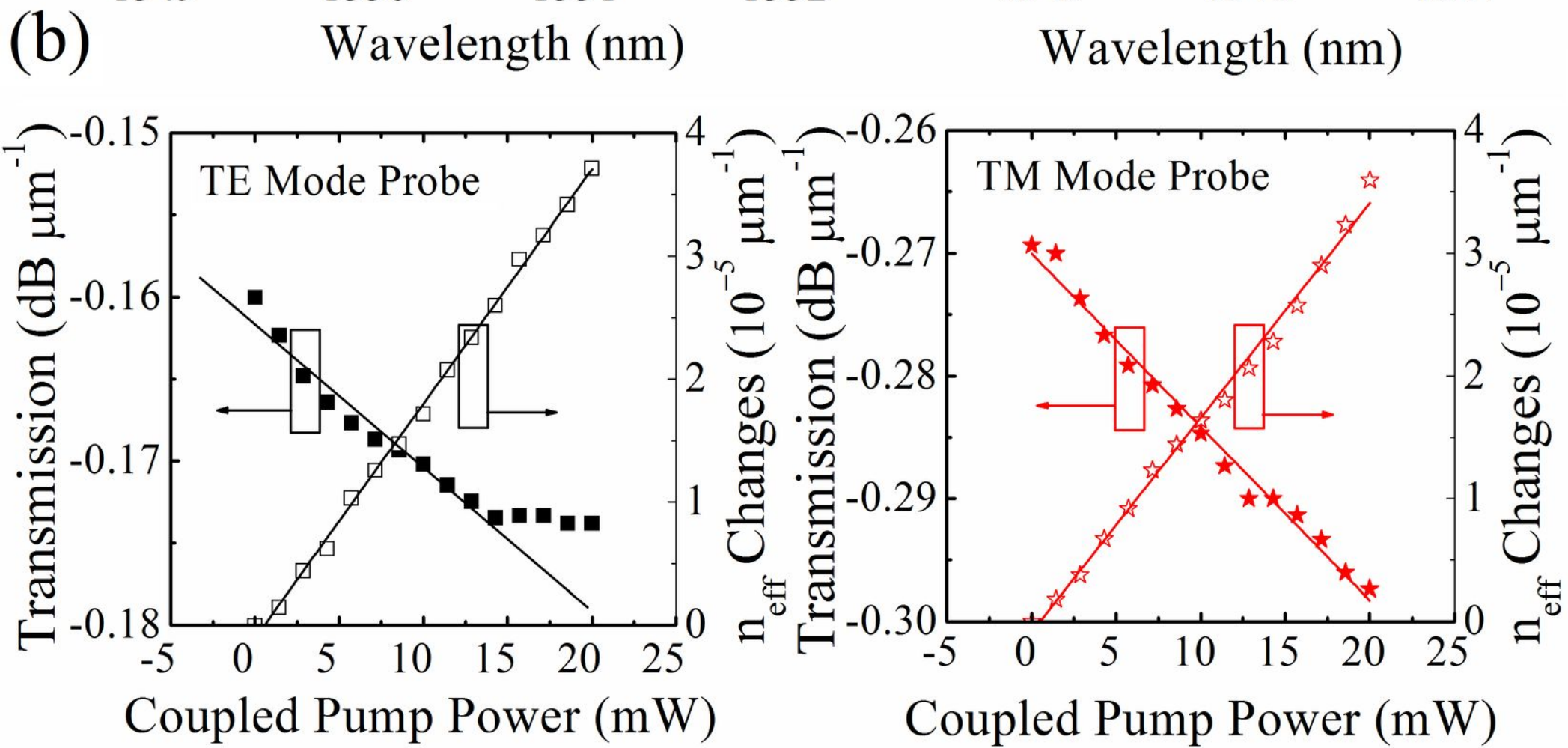